\documentclass[aps,prc,showpacs,twocolumn,superscriptaddress]{revtex4}
\usepackage{CJK}
\usepackage{multirow}
\usepackage{graphicx}
\usepackage{color}
\usepackage{amsmath}
\usepackage{amssymb}
\usepackage[colorlinks=true]{hyperref}

\definecolor{dgreen}{cmyk}{1.,0.,1.,0.4}        
\definecolor{orange}{cmyk}{0.,0.353,1.,0.}    

\begin{document}
\begin{CJK*}{GB}{} 
\title{
\begin{flushright}
{
}
\end{flushright}
Anisotropic distributions in a multi-phase transport model}

\author{You~Zhou}
\email{you.zhou@cern.ch}
\affiliation{Niels Bohr Institute, University of Copenhagen, Blegdamsvej 17, 2100 Copenhagen, Denmark}

\author{ Kai~Xiao}
\affiliation{Key Laboratory of Quark and Lepton Physics (MOE) and Institute of Particle Physics, Central China Normal University, Wuhan 430079, China}
\affiliation{South-Central University for Nationalities, Wuhan 430074, China}

\author{Zhao~Feng}
\affiliation{Key Laboratory of Quark and Lepton Physics (MOE) and Institute of Particle Physics, Central China Normal University, Wuhan 430079, China}

\author{Feng~Liu}
\email{fliu@mail.ccnu.edu.cn}
\affiliation{Key Laboratory of Quark and Lepton Physics (MOE) and Institute of Particle Physics, Central China Normal University, Wuhan 430079, China}

\author{ Raimond~Snellings}
\affiliation{Utrecht University,  P.O.Box 80000, 3508 TA Utrecht, The Netherlands}

\date{\today}

\begin{abstract}

With A Multi-Phase Transport (AMPT) model we investigate the relation between the magnitude, fluctuations and correlations of the initial state spatial anisotropy $\varepsilon_{n}$ and the final state anisotropic flow coefficients $v_{n}$ in Au+Au collisions at $\sqrt{s_{_{\rm NN}}}=$  200 GeV. 
It is found that the relative eccentricity fluctuations in AMPT account for the observed elliptic flow fluctuations, both are in agreement with the elliptic flow fluctuation measurements from the STAR collaboration. 
In addition, the studies based on 2- and multi-particle correlations and event-by-event distributions of the anisotropies suggest that the Elliptic-Power function is a promising candidate of the underlying probability density function of the event-by-event distributions of $\varepsilon_{n}$ as well as $v_{n}$. Furthermore, the correlations between different order symmetry planes and harmonics in the initial coordinate space and final state momentum space are presented. Non-zero values of these correlations have been observed. The comparison between our calculations and data will, in the future, shed new insight into the nature of the fluctuations of the Quark-Gluon Plasma produced in heavy ion collisions.

\end{abstract}


\pacs{25.75.Ld, 25.75.Gz}

\maketitle
\end{CJK*}

\clearpage

\section{Introduction}
\label{section1}

One of the fundamental questions in the phenomenology of Quantum Chromo Dynamics (QCD) 
is what the properties of matter are at the extreme densities and temperatures where quarks
and gluons are in a new state of matter, the so-called Quark Gluon Plasma (QGP).
Collisions of high-energy heavy-ions, at the Brookhaven Relativistic Heavy Ion Collider (RHIC) 
and the CERN Large Hadron Collider (LHC), allow us to create and study the properties of such a 
system in the laboratory.

The azimuthal anisotropy in particle production is, at these energies, an observable 
which provides experimental information on the Equation of State and the transport properties of the QGP.
This anisotropy is usually characterized by the Fourier flow-coefficients~\cite{Voloshin:1994mz},
\begin{eqnarray}
v_{n} &=& \langle\cos[n(\varphi-\Psi_{n})\rangle, 
\;\;\; {\rm or\ equivalently} \nonumber \\ 
v_n  &=& \langle e^{in\varphi}  e^{-in \Psi_n} \rangle,
\label{eq1}
\end{eqnarray}
where $\varphi$ is the azimuthal angle of the particles, $\Psi_{n}$ is the $n^{\rm{th}}$-order flow plane (or named final state symmetry plane) angle
and $\langle~\rangle$ denotes an average over the selected particles and events. 

In the last decade, the elliptic flow $v_{2}$~\cite{Alt:2003ab,Ackermann:2000tr,Adcox:2002ms,Back:2002gz,Aamodt:2010pa,ATLAS:2011ah,Chatrchyan:2012wg}, 
which is considered to correspond to the elliptical shape of the spatial overlap region in the system created 
in the collisions~\cite{Ollitrault:1992bk}, has received a lot of experimental and theoretical attention.
For a recent summary see~\cite{Voloshin:2008dg, Snellings:2011sz, Konchakovski:2012tm, Luzum:2013yya, Snellings:2014kwa,Jia:2014jca}. 
More recently, higher odd and even anisotropic flow coefficients are found to be also very 
important~\cite{Alver:2010gr}. 
Hydrodynamic calculations predict that these higher harmonics, 
such as the triangular flow $v_{3}$, are more sensitive to the shear viscosity to 
entropy density ratio $\eta/s$ of the QGP than $v_{2}$~\cite{Alver:2010dn}.
Furthermore, it is realized that the correlations between the symmetry planes and flow harmonics are sensitive to both the initial state and $\eta/s$~\cite{Qiu:2012uy, Deng:2011at, Niemi:2015qia}. The combined analysis of both $\varepsilon_{n}$ and $v_{n}$ distributions for a single harmonic, and the correlations between different orders of symmetry planes (direction of the flow vector) and flow harmonics (magnitude of the flow vector) could yield powerful constraints on both initial conditions and properties of the QGP. However, so far the investigation of the above mentioned combined analysis using a transport model was still lacking.

In this paper, we present the calculations of initial state anisotropies and final state anisotropic flow in Au+Au collisions at $\sqrt{s_{_{\rm NN}}} = 200$~GeV using the AMPT model~\cite{Lin:2004en}. 
We also investigate the relation between the magnitude, fluctuations and correlations of the initial state spatial eccentricity and final state anisotropic flow coefficients.
In addition, the correlations between different order symmetry planes and harmonics will be investigated in both initial  and final state, which will help us understand how they are modified during the expansion of the system.

\section{\bf A Multi--Phase Transport model}

A Multi-Phase Transport (AMPT) model~\cite{Lin:2004en} with a so-called string melting 
scenario has been used for these studies. 
The model consists of four main stages: initial conditions, partonic interactions, 
hadronization and finally hadronic rescattering. 

The initial conditions, which include the spatial and momentum distributions of minijet partons and 
soft string excitations, are obtained from the {\sc HIJING} model~\cite{Wang:2000bf}. 
The strings are converted into partons and the next stage, which models the interactions between all the partons, 
is based on {\sc ZPC}~\cite{Zhang:1997ej}.
The partonic cascade model {\sc ZPC} presently includes only 2-body processes with cross sections obtained 
from pQCD with screening masses. 
In {\sc ZPC}, the default value of the cross section is 3 mb. 
The transition from partonic- to hadronic-matter is modeled by a
simple coalescence model, which combines two quarks into mesons and
three quarks into baryons~\cite{Chen:2005mr}. 
Finally, to describe the dynamics of the subsequent hadronic stage, a hadronic cascade based on 
the {\sc ART} model~\cite{Li:1995pra} is used.
In this analysis, we used the default input parameters of AMPT with string melting suggested in~\cite{Lin:2004en}. The results are presented as a function of centrality, which determined by impact parameter $\bf {b}$, as used in~\cite{Xu:2011fe, Abelev:2013csa}. The possible effects of the fluctuations of impact parameters will be discussed in next section.

\section{Analysis Method and Definitions}

In this paper the anisotropic flow is calculated using the 2- and multi-particle 
cumulants method~\cite{Borghini:2001vi,Bilandzic:2010jr}, which was widely used at RHIC~\cite{Agakishiev:2011eq} and at the LHC~\cite{Aamodt:2010pa, ALICE:2011ab}. 
In this method, both 2- and multi-particle azimuthal correlations are analytically expressed 
in terms of a Q-vector, which is defined as:
 \begin{equation}
Q_{n} = \sum_{i=1}^{M}e^{in\varphi_{i}},
\end{equation}
where $M$ is the multiplicity of the selected particles and $\varphi$ is their azimuthal angle.

The single-event average 2-particle azimuthal correlations can be calculated via:
\begin{equation}
\langle 2 \rangle  = \frac{|Q_{n}|^{2} - M} {M(M-1)}.
\label{mean2}
\end{equation}
From this the event averaged 2-particle correlations, and the 2-particle cumulants can be obtained using:
\begin{equation}
c_{n}\{2\} \equiv \langle \langle 2 \rangle \rangle  = \frac{\sum_{events} (W_{\langle 2 \rangle})_{i} \langle 2 \rangle_{i} }{ \sum_{events} (W_{\langle 2 \rangle})_{i} },
\label{EqMeanMean2}
\end{equation}
where $W_{\langle 2 \rangle}$ is the event weight.
To minimize the effect of the varying multiplicity in certain centrality class determined by $\bf b$, we use the number of particle pairs as the event weight proposed in~\cite{Ante-Thesis}:
\begin{equation}
W_{\langle 2 \rangle} \equiv M(M-1)
\label{Eq:Weight2}
\end{equation}
The anisotropic flow from 2-particle cumulants, denoted as $v_{n}\{2\}$, is finally obtained from:
\begin{equation}
v_{n}\{2\} = \sqrt{c_{n}\{2\}}.
\label{Eq:v22NoGap}
\end{equation}

Unfortunately, the $v_{n}\{2\}$ contains contributions from so-called non-flow 
effects, which are additional azimuthal correlations not associated with the common symmetry planes, e.g.  
resonance decays, jet fragmentation, and Bose-Einstein correlations. 
They can be suppressed by appropriate kinematic cuts and therefore one can
introduce a gap in pseudorapidity between the particles used in the 2-particle Q-cumulant method~\cite{Zhou:2014bba}. 
For this we divide the whole event into two sub-events, $A$ and $B$, which are separated by a pseudorapidity gap $|\Delta\eta|$.
This modifies Eq.~(\ref{mean2}) to:
\begin{equation}
\langle 2 \rangle _{\Delta \eta} = \frac{Q_{n}^{A} \cdot Q_{n}^{B *}} {M_{A} \cdot M_{B}}, 
\label{Eqmean2Gap}
\end{equation}
where $Q_{n}^{A}$ and $Q_{n}^{B}$ are the flow vectors from sub-event $A$ and $B$, with $M_{A}$ and $M_{B}$ 
the corresponding multiplicities. 
The event weight from Eq.~(\ref{Eq:Weight2}) in this case becomes:
\begin{equation}
W_{\langle 2 \rangle _{\Delta\eta}} \equiv M_{A} \cdot M_{B}.
\label{Eq:Weight2etagap}
\end{equation}
Finally, inserting Eqs.~(\ref{Eqmean2Gap}) and (\ref{Eq:Weight2etagap}) into Eq.~(\ref{EqMeanMean2}), the $v_{n}$ from a 2-particle cumulant with a $\Delta\eta$ gap is given by:
\begin{equation}
v_{n}\{2, |\Delta\eta|\} = \sqrt{\langle \langle 2 \rangle \rangle _{\Delta\eta} }.
\label{Eq:v22Gap10}
\end{equation}

Instead of using kinematic cuts, the collective nature of anisotropic flow itself can be exploited to suppress 
non-flow contributions. 
Using multi-particle instead of 2-particle cumulants the aforementioned non-flow effects are strongly suppressed and 
no additional kinematic cuts are required. The $v_n$ calculated using cumulants are denoted as $v_n\{k\}$, were 
$k$ is $2,4,6\dots, m$ for the $m$-particle cumulant. Following the Q-cumulant method~\cite{Bilandzic:2010jr}, the single event average 4- and 6-particle correlations can be calculated as:
\begin{equation}
\begin{aligned}
\langle 4 \rangle & = [ \, |Q_{n}|^{4} + |Q_{2n}|^{2} - 2 \cdot {\rm{Re}} \left( Q_{2n}Q_{n}^{*}Q_{n}^{*} \right) \\
&~~  - 2 [ 2(M-2) \cdot |Q_{n}|^{2} - M(M-3) ] \, ]\\
& ~~ / [ M(M-1)(M-2)(M-3) ], \\
\langle 6 \rangle & =  [  \, |Q_{n}|^{6}  + 9 |Q_{2n}|^{2} |Q_{n}|^{2} - 6 \cdot {\rm{Re}} \left( Q_{2n} Q_{n} Q_{n}^{*} Q_{n}^{*} Q_{n}^{*} \right)  \\
& ~~ +4 \cdot {\rm{Re}} \left(  Q_{3n} Q_{n}^{*} Q_{n}^{*} Q_{n}^{*}  \right) - 12  \cdot {\rm{Re}}  \left( Q_{3n} Q_{2n}^{*} Q_{n}^{*} \right)  \\
&~~ +18 (M-4)  \cdot {\rm{Re}} \left(  Q_{2n} Q_{n}^{*} Q_{n}^{*} \right) + 4 |Q_{3n}|^{2} \\
&~~ - 9 (M-4) (|Q_{n}^{4}| + |Q_{2n}|^{2}) + 18 (M-2) (M-5) |Q_{n}|^{2} \\
&~~ - 6M(M-4)(M-5) \, ] \\
& ~~/ \left[ M(M-1)(M-2)(M-3)(M-4)(M-5) \right]
\end{aligned}
\end{equation}

Then the multi-particle cumulants are obtained from:
\begin{equation}
\begin{aligned}
c_{n}\{4\} & = \langle \langle 4 \rangle \rangle - 2 \, \langle \langle 2 \rangle \rangle^{2},\\
c_{n}\{6\} & = \langle \langle 6 \rangle \rangle - 9 \, \langle \langle 2 \rangle \rangle \cdot \langle \langle 4 \rangle \rangle + 12 \, \langle \langle 2 \rangle \rangle^{3},
\end{aligned}
\label{Eq:c46}
\end{equation}
here $\langle\langle~\rangle\rangle$ denotes the average over all particles over all events.
In the end the $v_{n}$ from 4- and 6-particle cumulant, denoted as $v_{n}\{4\}$ and $v_{n}\{6\}$, are  obtained using:
\begin{equation}
\begin{aligned}
v_{n}\{4\} & = \sqrt[4] {- \, c_{n}\{4\}},\\
v_{n}\{6\} & = \sqrt[6]{c_{n}\{6\}} .
\end{aligned}
\label{Eq:c46}
\end{equation}

The 2- and multi-particle cumulants have different contributions from flow fluctuations. 
The contribution from flow fluctuations is positive for the 2-particle cumulant and negative for 
the multi-particle cumulant~\cite{Miller:2003kd, Zhu:2005qa}. 
When non-flow effects are negligible for the 2-particle cumulant, and 
if $\sigma_{v_n} \ll \overline{v}_n$, the $v_n$ from the cumulants are up to order $\sigma^2_{v_n}$ given by:
\begin{equation}
\begin{aligned}
v_{n}\{2\}^{2} \approx & ~ \overline{v}_{n}^2 + \sigma_{v_{n}}^{2}, \\
v_{n}\{4\}^{2} \approx & ~ \overline{v}_{n}^2 - \sigma_{v_{n}}^{2}, \\
v_{n}\{6\}^{2} \approx & ~ \overline{v}_{n}^2 - \sigma_{v_{n}}^{2}, 
\end{aligned}
\label{Eq:v2246MeanV}
\end{equation}
where $\overline{v}_{n}$ and $\sigma_{v_{n}}$ are the mean and standard deviation of the $v_{n}$ 
distributions, respectively. 
In the special case that the underlying probability density function ($p.d.f.$) of $v_n$ is described by a Bessel-Gaussian function, Eq.~(\ref{Eq:v2246MeanV}) is an exact solution, independent of the magnitude of the flow fluctuations.
This shows that for a Bessel-Gaussian $p.d.f.$ all the multi-particle cumulants will be identical.

It is thought that the development of anisotropic flow is controlled by the anisotropies in the 
pressure gradients which in turn depend on the shape and structure of the initial density profile. 
The latter can be characterized, in analogy with the flow Fourier coefficients and flow angles of Eq.~(\ref{eq1}),  
by a set of harmonic anisotropy coefficients $\varepsilon_n$ and associated symmetry angles $\Phi_n$ of the initial 
spatial distribution:
\begin{equation}
\varepsilon_n e^{in\Phi_n} \equiv - \frac{\int  r \, dr \, d\phi \, r^n e^{in\phi} \, e(r,\phi)}
                                                              {\int r \, dr \, d\phi \, r^n e(r,\phi)}\  (n>1),
\label{eq3.1}
\end{equation}
where $e(r,\phi)$ is the initial energy density distribution in the plane transverse to the beam direction.
Also these anisotropy coefficients can be calculated using cumulants and the same relation between 
$\overline{\varepsilon}_{n}$ and $\sigma_{\varepsilon_{n}}$ follows as in Eq.~\ref{Eq:v2246MeanV}, 
replacing $v_n$ with $\varepsilon_n$.

\section{Results and Discussion}

\begin{figure}[h]
 \begin{center}
   \includegraphics[width=0.48\textwidth]{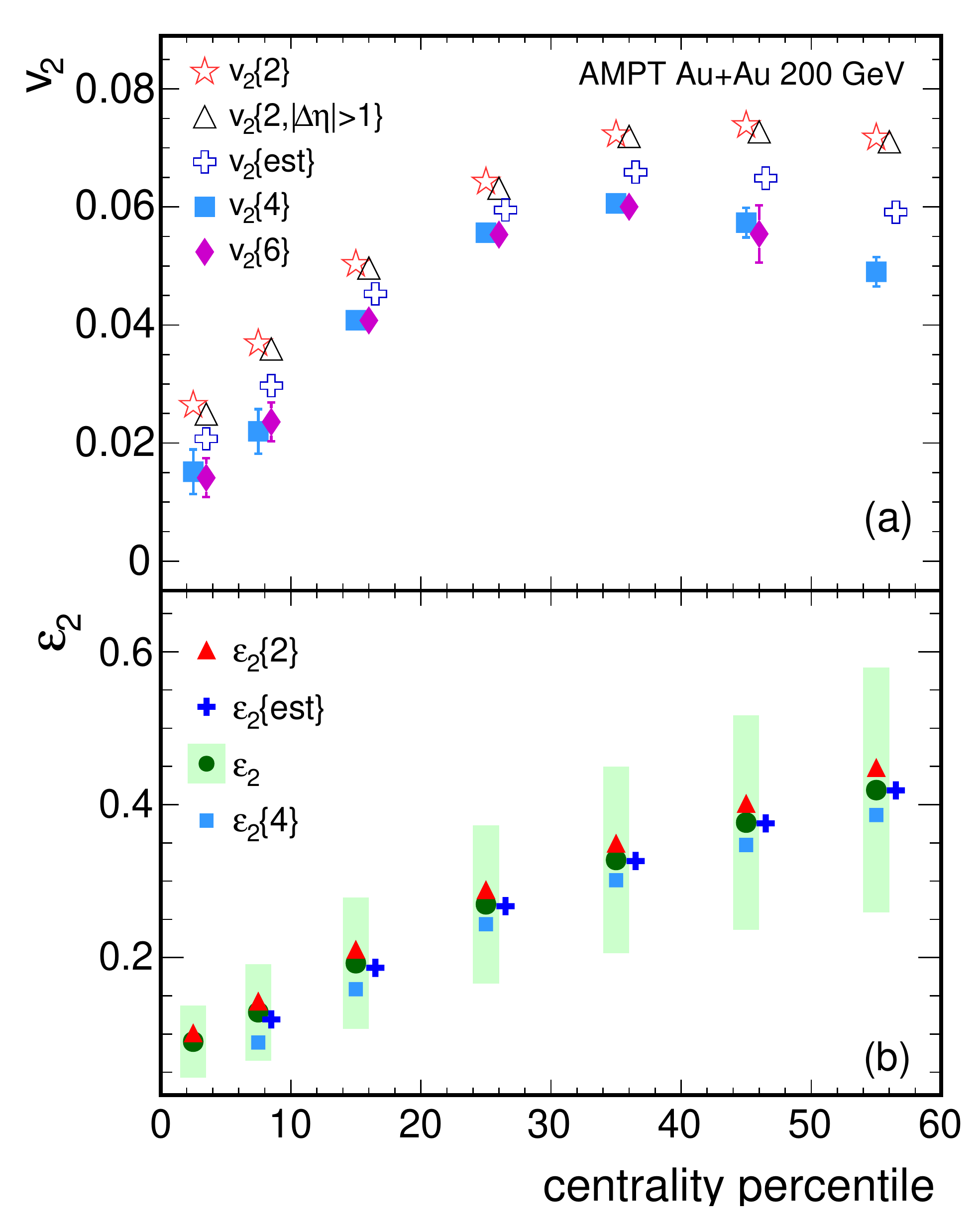}
   \caption{(Color online) Centrality dependence of (a) $v_{2}$ and (b) $\varepsilon_{2}$ 
    in Au+Au collisions at  $\sqrt{s_{_{\rm NN}}}=200$ GeV in AMPT. 
    For the definition of the symbols see text.
    }
    \label{fig:1}
 \end{center}
\end{figure}

The centrality dependence of $v_{2}$ from the AMPT model calculations is shown in Fig.~\ref{fig:1}a. 
For the 2-particle cumulant calculations, we plot both $v_{2}\{2\}$ from Eq.~(\ref{Eq:v22NoGap}) 
and $v_{2}\{2,|\Delta\eta|>1\}$ from Eq.~(\ref{Eq:v22Gap10}). 
It is seen that the $v_{2}\{2,|\Delta\eta|>1\}$ is compatible with $v_{2}\{2\}$, which indicates that short range non-flow contributions
are very small in AMPT. 
Figure~\ref{fig:1}a also shows the calculated $v_{2}\{4\}$ and $v_{2}\{6\}$. 
Because the non-flow contributions to the integrated flow are very small, the observed significant difference between 
$v_{2}\{2\}$ and $v_{2}\{4\}$ is due to elliptic flow fluctuations. 
In addition we observe that there is very good agreement between $v_{2}\{4\}$ and $v_{2}\{6\}$. This seems to agree with the expectation that the underlying $p.d.f.$ is a Bessel-Gaussian function, which predicts $v_{2}\{4\} = v_{2}\{6\}$.
However, it is still important to point it out that in fact $v_{2}\{4\} \approx v_{2}\{6\}$ is valid irrespectively of the details of the underlying model of flow fluctuations, under the assumption $\sigma_{v_{n}} \ll \overline{ v_{n}}$~\cite{Ante-Thesis}.

From $v_{2}\{2,|\Delta\eta|>1\}$ and $v_{2}\{4\}$, we can estimate $\overline{v}_{2}$ in Eq.~(\ref{Eq:v2246MeanV}) using:
\begin{equation}
\overline{v}_{2}\{\mathrm{est}\} = \sqrt {\frac{v_{2}\{2,|\Delta\eta|>1\}^{2} + v_{2}\{4\}^{2}}{2}}.
\end{equation}

Figure~\ref{fig:1}b shows $\varepsilon_{2}$ as function of centrality. 
Here $\varepsilon_{2}$ can be calculated from the initial spatial parton distributions in the AMPT model. 
The relative fluctuation of $\varepsilon_{2}$, named $\sigma_{\varepsilon_{2}}$, is shown by the green band as the uncertainty of $\overline{\varepsilon}_{2}$. 
The $\varepsilon_{2}\{2\}$ and $\varepsilon_{2}\{4\}$ correspond to the 2- and 4-particle cumulant 
definition but are evaluated using the initial spatial coordinates.
We see that $\varepsilon_{2}\{2\}$, $\varepsilon_{2}$ and $\varepsilon_{2}\{4\}$ increase monotonically 
up to 60\% centrality percentile. In contrast, the $v_2$, plotted in Fig.~\ref{fig:1}a, 
starts to saturate or decrease from 40\% centrality percentile. 
This difference between the centrality dependence of $v_2$ and $\varepsilon_2$ is generally understood to be due to 
the fact that the efficiency of converting the initial eccentricity into final elliptic flow decreases towards 
peripheral collisions because of the decreasing number of reinteractions in a smaller system.

\begin{figure}[thb]
 \begin{center}
   \includegraphics[width=0.48\textwidth]{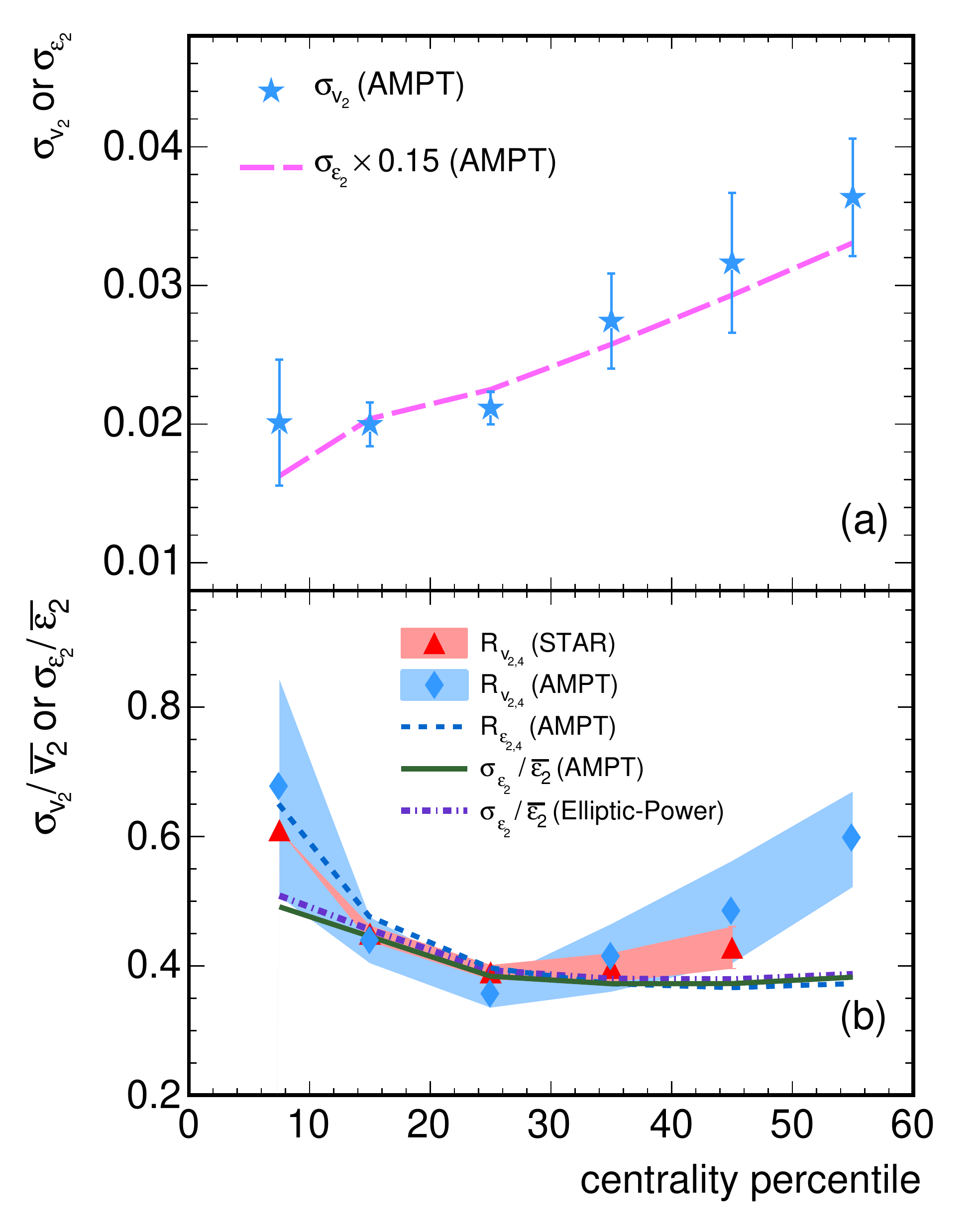}
   \caption{(Color online) Centrality dependence of (a) elliptic flow fluctuation and (b) relative elliptic flow (eccentricity) fluctuations in Au+Au collisions at 200 GeV in AMPT StringMelting. 
   (a) The full stars are the estimated elliptic flow fluctuation $\sigma_{v_{2}}$ and the dashed line is the relative eccentricity fluctuations scaled by a factor of 0.15; 
   (b) the estimated relative elliptic flow fluctuation $R_{v_{2,4}}$, the relative eccentricity and true eccentricity fluctuation for the AMPT model are shown using full diamonds, a dashed line and solid line. 
   In addition, the measurement from the STAR Collaboration are plotted using full triangles and the shadow shows its uncertainty~\cite{Agakishiev:2011eq}. 
   }
  \label{fig:2}
 \end{center}
\end{figure}

From the calculated anisotropic flow, we can investigate the relation between the fluctuations in $\varepsilon_n$ and $v_n$. 
In Fig.~\ref{fig:2}a we plot the centrality dependence of the elliptic flow fluctuations
$\sigma_{v_{2}}$ estimated via 
\begin{equation}
 \sigma_{v_{2}} = \sqrt{\frac{v_{2}\{2,|\Delta\eta|>1\}^{2}-v_{2}\{4\}^{2}}{2}}.
\label{Eq:sigmav2}
\end{equation}
We find that $\sigma_{v_{2}}$ increases toward peripheral collisions and its magnitude is significant compared to the magnitude of $v_2$. 
Assuming $v_{2} \propto \varepsilon_{2}$ and using Eq.~(\ref{Eq:v2246MeanV}), we also have  $v_{2}\{2\} \propto \varepsilon_{2}\{2\}$ and $\sigma_{v_{2}} \propto \sigma_{\varepsilon_{2}}$. Hence, figure~\ref{fig:2}a also shows the scaled $\sigma_{\varepsilon_{2}}$ calculated as $\sigma_{\varepsilon_{2}} = \sqrt{(\varepsilon_{2}\{2\}^{2}-\varepsilon_{2}\{4\}^{2})/2}$. It shows that the eccentricity fluctuations describe the centrality dependence of the elliptic flow fluctuations quite well, which might indicates that they are the dominant contribution for the observed elliptic flow fluctuations.  

The estimated relative flow fluctuations, can be calculated using:
\begin{equation}
R_{v_{_{2,4}}}= \sqrt{ \frac{v_{2}\{2,|\Delta\eta|>1\}^{2}-v_{2}\{4\}^{2}}{v_{2}\{2,|\Delta\eta|>1\}^{2}+v_{2}\{4\}^{2}}} \approx \frac{{\sigma_{v}}_{2}}{\overline{v}_{2}},
\label{Eq:R24}
\end{equation}
and its centrality dependence is shown in Fig.~\ref{fig:2}b. 
The magnitude of the relative fluctuations ranges from 0.4 to 0.6 in central to mid-peripheral collisions.
This shows that the assumption $\sigma_{v_2} \ll \overline{v}_2$  
does not hold for these centralities.

The estimated relative eccentricity fluctuations $R_{\varepsilon_{_{2,4}}}$ 
can be calculated analogously to the $R_{v_{_{2,4}}}$: 
\begin{equation}
R_{\varepsilon_{_{2,4}}}= \sqrt{ \frac{\varepsilon_{2}\{2\}^{2}-\varepsilon_{2}\{4\}^{2}}
{\varepsilon_{2}\{2\}^{2}+\varepsilon_{2}\{4\}^{2}}} \approx \frac{{\sigma_{\varepsilon}}_{2}}{\overline{\varepsilon}_{2}},
\end{equation}
and compared to $R_{v_{_{2,4}}}$. This comparison does not depend on the assumption that 
the relative fluctuations are small or the underlying $p.d.f.$ is a Bessel-Gaussian.
We notice that the $R_{\varepsilon_{_{2,4}}}$ (also plotted in Fig.~\ref{fig:2}b) is compatible
with $R_{v_{_{2,4}}}$ for central and mid-central collisions. 
To test if this is only accidental we also study the $v_2$ in AMPT with an approximately three times larger partonic cross section (10 mb).
The magnitudes of both $v_{2}\{2\}$ and $v_{2}\{4\}$ increase significantly, 
however, the consistency between $R_{\varepsilon_{2,4}}$ and $R_{v_{2,4}}$ continues to hold. 
This is expected if the relative elliptic flow fluctuations depend only on the eccentricity fluctuations, which again shows that eccentricity fluctuations play an important role in the development of elliptic flow fluctuations in the AMPT model.

In addition, we calculate directly from the initial state of the AMPT model 
the true relative eccentricity fluctuations 
$\sigma_{\varepsilon_{2}}/\overline{\varepsilon}_{2}$ (plotted as the the green solid line in Fig.~\ref{fig:2}b). 
The results are consistent with the estimated relative eccentricity fluctuations in the 20--50\% centrality percentile, 
while they deviate for central and peripheral collisions.  
Because the assumption $\sigma_{\varepsilon_{2}} \ll \varepsilon_{2}$ is not satisfied over the whole centrality range this indicates 
that for 20--50\% the fluctuations might be approximately described by a Bessel-Gaussian but to describe the overall event-by-event $v_{2}$ distributions, we need to search for a better candidate of the underlying $p.d.f.$
To see how these fluctuations compare to experimental data we compare to STAR 
measurements~\cite{Agakishiev:2011eq}.  The relative elliptic flow fluctuations measured 
in Au+Au collisions at 200 GeV are plotted in Fig.~\ref{fig:2}b 
and are in very good agreement with the AMPT model calculations.


Hydrodynamic calculations have shown that in a given event, the $v_{n}$ is a linear response to the initial anisotropy ($v_{n} = k_{n} \,\varepsilon_{n}$), for $n=$ 2, 3~\cite{Niemi:2012aj}.
In the discussion above, we found that the elliptic flow and its fluctuations can be nicely described by the initial eccentricity together with its fluctuations. In order to further constrain the underlying $p.d.f.$ of the $v_{n}$ and $\varepsilon_{n}$ distributions, we study the event-by-event distributions of $\varepsilon_{n}$, simulated by AMPT. 

In this study, the event-by-event $\varepsilon_{n}$ distributions are investigated in the selected centrality classes. To better extract the information on the underlying $p.d.f.$ of the $\varepsilon_{n}$ distributions, several candidate functions are used in this paper. 
One popular parameterization of the $\varepsilon_{n}$ distribution is the Bessel-Gaussian distributions~\cite{Voloshin:2007pc}:
\begin{equation}
p(\varepsilon_{n}) = \frac{\varepsilon_{n}}{\sigma^{2}} I_{0}\left(  \frac{\varepsilon_{n} \varepsilon_{n}}{\sigma^{2}} \right) {\rm exp} \left( - \frac{ \varepsilon_{0}^{2} + \varepsilon_{n}^{2}} {2\sigma^{2}}  \right) ,
\end{equation}
where $\varepsilon_{0}$ is the anisotropy $w.r.t.$ the reaction plane and $\sigma$ is the fluctuation in the spatial anisotropy. 
It was already shown in this paper as well as in previous studies~\cite{Voloshin:2007pc, Broniowski:2007ft} that a Bessel-Gaussian distribution nicely describes the $\varepsilon_{2}$ distributions for mid-central collisions. However, it is not expected to work perfectly in peripheral collisions, due to the lack of a constraint that $\varepsilon_{2} <$ 1 in each event~\cite{Yan:2014afa}.
To fix this problem, a simple one-parameter power-law distribution~\cite{Yan:2013laa}
\begin{equation}
p(\varepsilon_{n}) = 2 \alpha \, \varepsilon_{n} \left( 1- \varepsilon_{n}^{2} \right)^{\alpha -1},
\end{equation}
was proposed to parameterized the fluctuations driven anisotropies. Here $\alpha$ quantifies the fluctuations. 

Recently, a new function, named ``Elliptic Power" distribution was proposed in~\cite{Yan:2014afa} as:
\begin{equation}
p(\varepsilon_{n}) = \frac{\alpha \, \varepsilon_{n}}{\pi}  \left( 1- \varepsilon_{0}^{2} \right)^{\alpha + \frac{1}{2}}  \int_0^{2\pi} \frac{ \left( 1- \varepsilon_{n}^{2} \right)^{\alpha - 1} \, \mathrm{d}\phi} { \left(  1- \varepsilon_{0} \,\varepsilon_{n} \cos \phi  \right)^{2\alpha +1}},
\end{equation}
where $\alpha$ and $\varepsilon_{0}$ have the same meaning as above.
In our paper, these three candidates of the underlying $p.d.f.$s are fitted to the $\varepsilon_{n}$ distributions in AMPT model.
 
 \begin{figure}[thb]
\begin{center}
\includegraphics[width=0.48\textwidth]{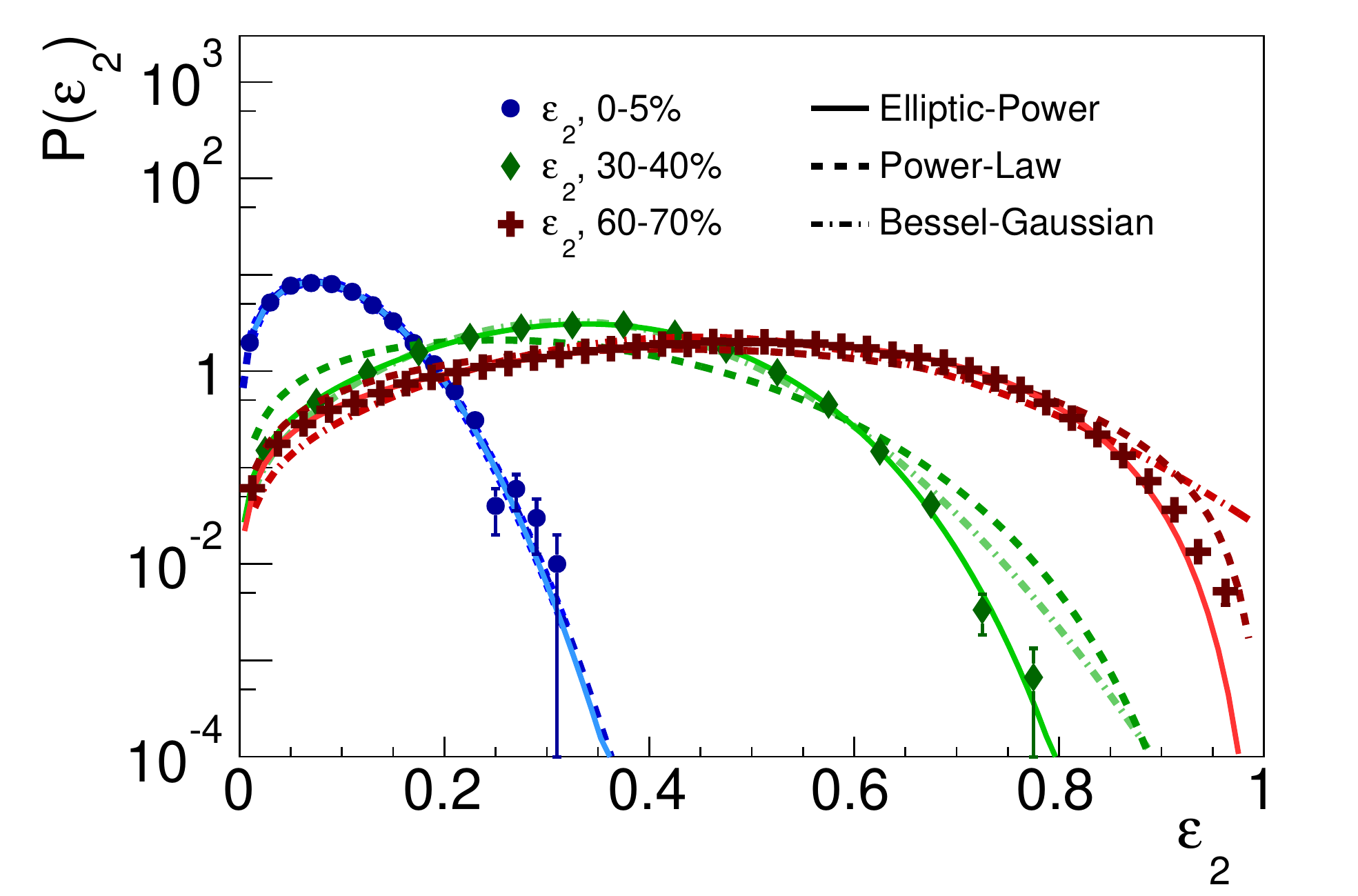}
\caption{(Color online) $\varepsilon_{2}$ distributions in the AMPT initial state}
\label{fig:e2dis}
\end{center}
\end{figure}
 
Figure~\ref{fig:e2dis} shows the $\varepsilon_{2}$ distributions in the centrality range 0-5\%, 30-40\% and 60-70\% from the AMPT initial state. We fit the three distributions with the Elliptic-Power (solid lines), Power-Law (dash line), and Bessel-Gaussian functions (dot-dash line). It is shown that in 0-5\% these three functions give consistent results and they all fit the $\varepsilon_{2}$ distributions quite well. 
It is understood that in the case where $\varepsilon_{0} \ll$ 1 (small anisotropy from the reaction plane) and where $\alpha >$ 1 (strong fluctuations), the Elliptic Power distribution turns into a Bessel-Gaussian distribution; while with $\varepsilon_{0} =$ 0 (anisotropy is solely due to fluctuations), the Elliptic Power distribution reduces to a Power-Law distribution. The nice agreement between the three functions shows that the $\varepsilon_{0}$ must be very small, which means that the eccentricity is generated mainly by fluctuations. In 30-40\%, it is clear that only the Elliptic-Power function agrees with the $\varepsilon_{2}$ distributions. For more peripheral collisions, the Elliptic-Power function describes the $\varepsilon_{2}$ distributions still quite well, while the Bessel-Gaussian can't quantitatively reproduce the $\varepsilon_{2}$ distributions. 
The Power-Law function, on the other hand, is not expected to describe the $\varepsilon_{2}$ distributions in non-central collisions, since the eccentricity from the reaction plane ($\varepsilon_{0}$) is missing and for non-central collisions $\varepsilon_{0}$ is non-zero and important to describe the $\varepsilon_{2}$ distributions. 
Furthermore, we calculate $\sigma_{\varepsilon_{2}}/\overline{\varepsilon}_{2}$ from the Elliptic-Power function with parameters extracted from the fits of the $\varepsilon_{2}$ distributions. The result shown in Fig.~\ref{fig:2}b is in nice agreement with the $\sigma_{\varepsilon_{2}}/\overline{\varepsilon}_{2}$ directly from the AMPT initial state.
For describing all the centralities the Elliptic-Power function gives the best description of the $\varepsilon_{2}$ distributions. Unfortunately, the expectation of a 1\% difference between the 4- and 6-particle cumulants ~\cite{Yan:2014afa} from the Elliptic-Power function can not be tested with the current statistics.

\begin{figure}[thb]
 \begin{center}
   \includegraphics[width=0.48\textwidth]{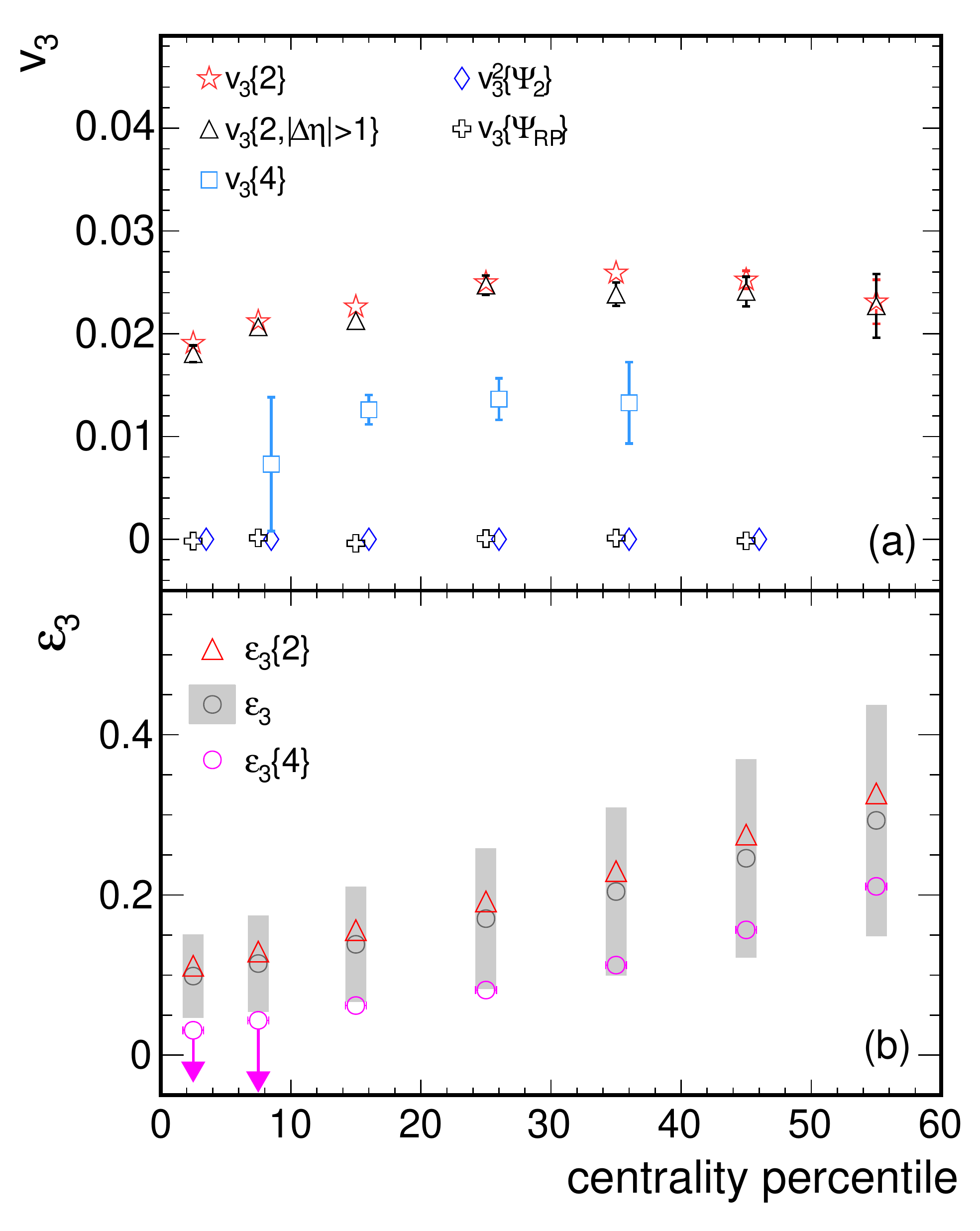}
   \caption{
   (Color online) 
   (a) Centrality dependence of $v_{3}$ in Au+Au collisions at 200 GeV from AMPT model calculations. The open stars (red), triangles (black), circles (gray), diamonds (blue) and crosses (black) 
   for $v_{3}\{2\}$, $v_{3}\{2,|\Delta\eta|>1\}$, $v_{3}^{2}\{\Psi_{2}\}$ and $v_{3}\{\Psi_{\rm RP}\}$ respectively. 
   (b) The corresponding $\varepsilon_3$ as a function of centrality.}
 \label{fig:3}
 \end{center}
\end{figure}

It is shown in hydrodynamic calculations that the higher harmonic flow coefficients are more sensitive to both the 
kinematic viscosity and to the initial geometry and its fluctuations~\cite{Alver:2010dn}.
In absence of fluctuations $v_3$ is zero due to symmetry constraints. Figure~\ref{fig:3}a shows the centrality dependence of $v_{3}$ obtained for AMPT events from different analysis methods. 
If $v_3$ is due to event-by-event fluctuations in the initial spatial density distributions we would expect that there is no, 
or only a very small, correlation between both the reaction plane angle $\Psi_{\mathrm{RP}}$ (spanned by the impact parameter vector and the beam direction) and the angle of the $2^{nd}$-order flow plane $\Psi_{2}$ with respect to the 3$^{rd}$-order flow plane $\Psi_{3}$.
The correlations between $\Psi_{3}$ and $\Psi_{\mathrm{RP}}$ can be studied via:
\begin{equation}
\begin{split}
v_{3}\{\Psi_{\rm RP}\} &= \langle \langle\cos 3(\varphi-\Psi_{\mathrm{RP}})\rangle \rangle \\
&= \langle \langle \cos3(\varphi-\Psi_{3}) \, \cos3(\Psi_{3} - \Psi_{\mathrm{RP}})\rangle  \rangle\\
&= \langle v_{3} \, \langle\cos 3(\Psi_{3} - \Psi_{\mathrm{RP}})\rangle \rangle.
\end{split}
\end{equation}
In Fig.~\ref{fig:3}a we observe that $v_{3}\{\Psi_{\rm RP}\}$ is consistent with zero.  
It indeed shows that there is no correlation (or an extremely weak correlation) 
between $\Psi_{3}$ and $\Psi_{\mathrm{RP}}$ in the presented centrality range.
We also see that both the $v_{3}\{2\}$ and $v_{3}\{4\}$ are nonzero and show only a weak centrality dependence, which is in qualitative agreement with earlier observations based on hydrodynamic model calculations~\cite{Alver:2010dn}. 
 
 \begin{figure}[thb]
\begin{center}
\includegraphics[width=0.45\textwidth]{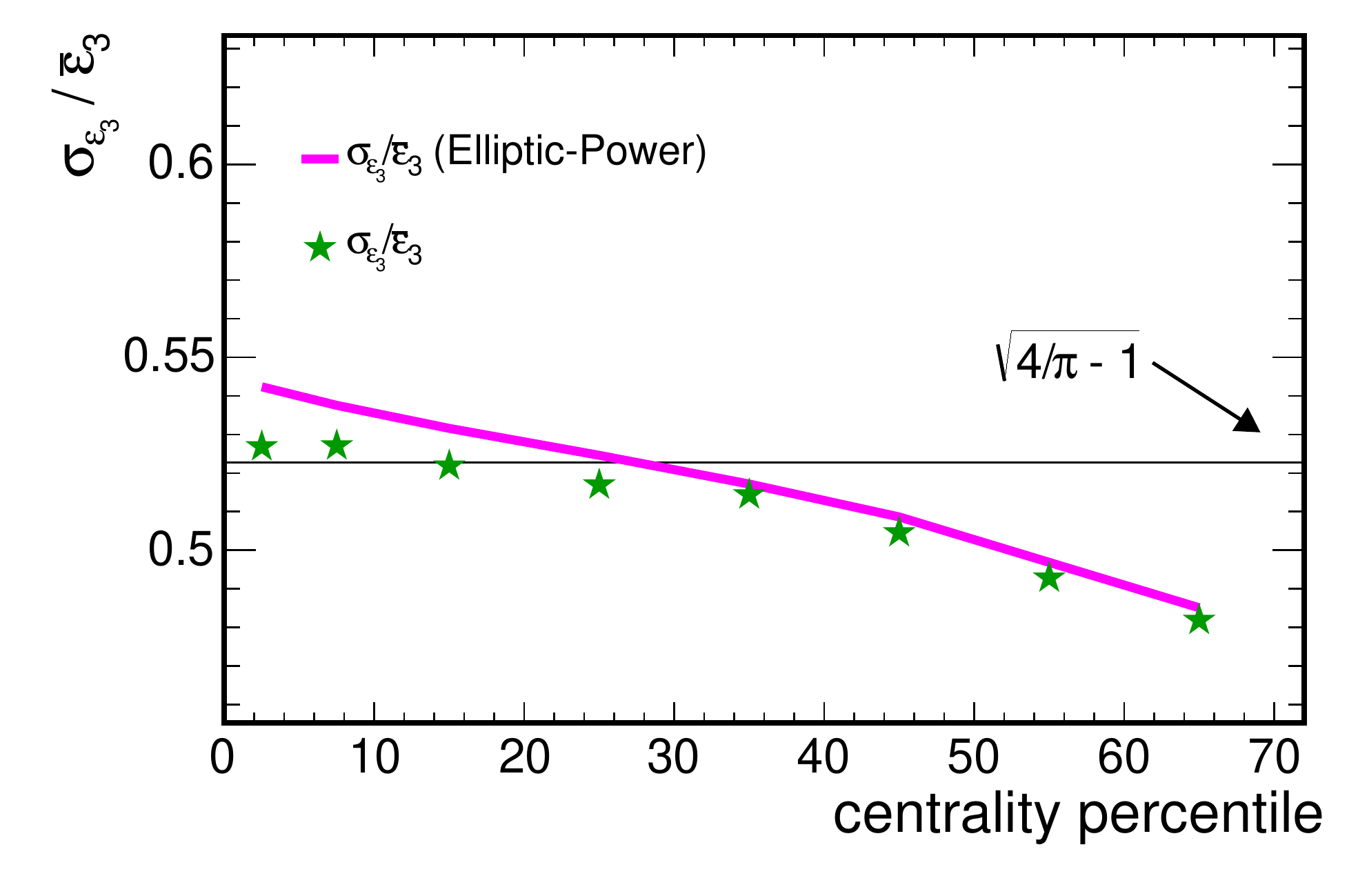}
\caption{(Color online) $\sigma_{\varepsilon_{3}}/\overline{\varepsilon}_{3}$ distributions in AMPT.}
\label{fig:sigmaOverE3}
\end{center}
\end{figure}
 
In Fig.~\ref{fig:3}b the corresponding $\varepsilon_{3}\{2\}$, $\varepsilon_{3}\{4\}$ and $\varepsilon_{3}$ are plotted which, in contrast with $v_{3}\{2\}$ and $v_{3}\{4\}$, increase with increasing centrality by a factor two, this difference might be due to strong viscous damping effects on $v_{3}$ compared to $v_{2}$~\cite{Alver:2010dn}.
The expectation that $\varepsilon_{3}\{4\} = \varepsilon_{3}\{\Psi_{\rm RP}\} =$ 0 if the $p.d.f.$ would be a Bessel-Gaussian function is not observed in the 5-40 \% centrality percentile range, as is shown in Fig.~\ref{fig:3}~\footnote{Due to large statistical fluctuations in most central collisions, the $\varepsilon_{3}^{2}\{4\} <$ 0, we therefore presented $\sqrt{-\varepsilon_{3}^{2}\{4\}}$ in most central collisions and the arrows show the range from $-\sqrt{-\varepsilon_{3}^{2}\{4\}}$ to $\sqrt{-\varepsilon_{3}^{2}\{4\}}$}. Similar non-zero $\varepsilon_{3}\{4\}$ were found for wider centrality ranges from both MC-Glauber and MC-KLN calculations~\cite{Zhou:2011np}. 
In fact, non-zero values of not only $v_{3}\{4\}$ but also $v_{3}\{6\}$ have been measured in experiments~\cite{Bilandzic:2012an}.
Compared to the Bessel-Gaussian distribution, which was the widely used description of the underlying $p.d.f$,  the Elliptic-Power function seems to gives an improved description of $\sigma_{\varepsilon_{3}}/\varepsilon_{3}$ and agrees with a non-zero value of multi-particle cumulants of $\varepsilon_{3}$ and $v_{3}$.

\begin{figure}[thb]
\begin{center}
\includegraphics[width=0.48\textwidth]{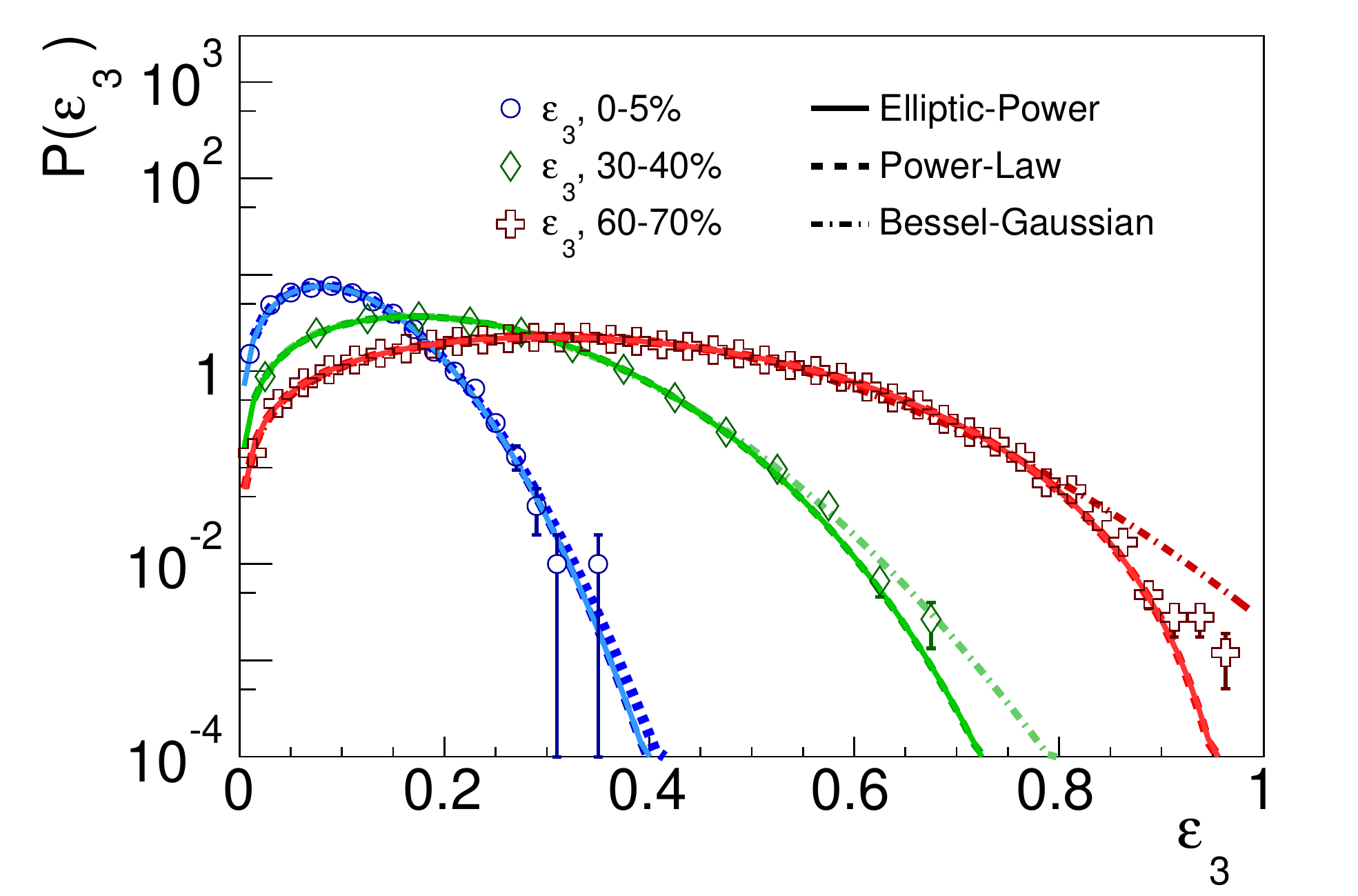}
\caption{(Color online) $\varepsilon_{3}$ distributions in the AMPT initial state}
\label{fig:e3dis}
\end{center}
\end{figure}

For completeness, Fig.~\ref{fig:e3dis} shows the $\varepsilon_{3}$ distributions in the centrality range 0-5\%, 30-40\% and 60-70\% from the AMPT initial state. Similar to the study of the $\varepsilon_{2}$ distributions, we fit all the $\varepsilon_{3}$ distributions with the Elliptic-Power (solid lines), Power-Law (dash line), and Bessel-Gaussian functions (dot-dash line).
It is seen that the Bessel-Gaussian function reproduces the $\varepsilon_{3}$ fairly well except for peripheral collisions. Nevertheless, the expectation of a Bessel-Gaussian, which is $\varepsilon_{3}\{4\} = \varepsilon_{3}\{6\} = \varepsilon_{3}\{\rm RP\} =0$ (as well as $v_{3}\{4\} = v_{3}\{6\} = v_{3}\{\rm RP\} =0$ ), does not agree with the non-zero $\varepsilon_{3}\{4\}$ and $v_{3}\{4\}$ presented above. The Bessel-Gaussian function is therefore not a candidate of the underlying $p.d.f.$ of $v_{n}$ and $\varepsilon_{n}$.
On the other hand, since triangularity is expected to be solely created by initial geometry fluctuations, its distributions should be well reproduced by a single-parameter Power-Law function~\cite{Yan:2014nsa}. Indeed nice agreements between the fits of the Power-Law and the $\varepsilon_{3}$ distributions are observed for the presented centralities. 
When testing the two parameter Elliptic-Power function, it turns out, as expected from the nice fit of the Power-Law function, that the parameter $\varepsilon_{0}$ (triangularity $w.r.t.$ reaction plane) is very close to 0. The nice descriptions of the $\varepsilon_{3}$ distributions by the Power-Law function and the Elliptic-Power function with parameter $\varepsilon_{0} \sim$ 0 confirms that the contributions of the reaction plane to $\varepsilon_{3}$ is very weak (or zero). This also agrees with our results of $v_{3}\{\rm RP\} =$ 0, displayed in Fig.~\ref{fig:3}, and with the experimental measurements of ALICE~\cite{ALICE:2011ab}. Furthermore, we notice that the true $\sigma_{\varepsilon_{3}} / \varepsilon_{3}$ from the AMPT initial state shows a decreasing trend from central to peripheral collisions. This decreasing trend is captured by the Elliptic-Power function quite well, and clearly disagrees with the expectation of a Bessel-Gaussian type $p.d.f.$,  which predicts a constant value of $\sqrt{4/\pi -1}$ for the entire centrality range~\cite{Broniowski:2007ft, Voloshin:2008dg}.
It is shown that the Elliptic-Power function gives a better description of the $\varepsilon_{2}$ and $\varepsilon_{3}$ distributions simultaneously, matches the multi-particle cumulants observables and the correlations $w.r.t.$ the reaction plane, and also matches the experimental measurements~\cite{ALICE:2011ab}. This shows that the Elliptic-Power distributions could be a promising candidate of the underlying $p.d.f.$ of $v_{n}$ (and $\varepsilon_{n}$), which helps us to better understand the initial geometry and its event-by-event fluctuations.

In addition, it is important to understand whether there is a correlation between different order flow vectors, including both the flow planes (direction of the flow vector) and the harmonics (magnitude of the flow vector), and how this correlation between flow vectors modifies the underlying $p.d.f.$
One example of the correlation between different flow planes is the correlation of $\Psi_{2}$ and $\Psi_{3}$ measured using a 5-particle cumulant:
\begin{equation}
v_{3}^{2}\{\Psi_{2}\}= \frac{\langle\cos(2\varphi_{1}+2\varphi_{2}+2\varphi_{3}-3\varphi_{4}-3\varphi_{5})\rangle}{v_{2}^{3}},
\end{equation}
\cite{Ante-Thesis} 
(also denoted by $v_{23}\equiv v\{2,2,2,-3,-3\}$ in~\cite{Bhalerao:2011yg}).
Figure~\ref{fig:3}a shows that $v_{3}^{2}\{\Psi_{2}\}$ from the AMPT model is consistent with zero, and since both $v_2$ and $v_3$ are non-zero, this result proves that there is no correlation between $\Psi_{2}$ and $\Psi_{3}$ in the AMPT calculations. 

\begin{figure}[thb]
 \begin{center}
   \includegraphics[width=0.48\textwidth]{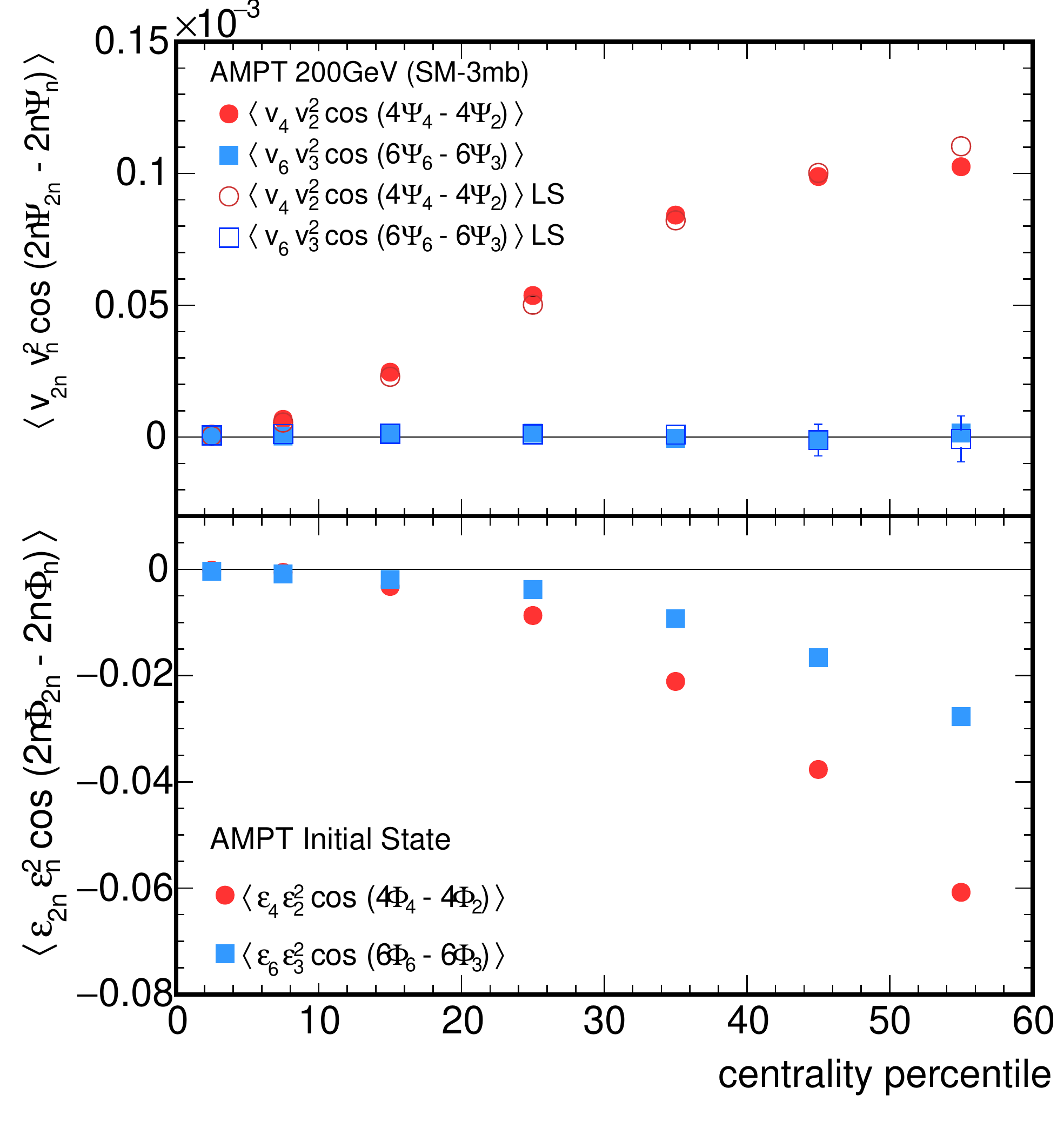}
   \caption{(Color online) Centrality dependence of the final state QC$\{3\}$(top) 
and the 2-plane correlations in the initial state (bottom) in Au+Au collisions at 200 GeV 
by AMPT StringMelting.  }
 \label{fig:7}
 \end{center}
\end{figure}

In addition to the correlations between the $2^{nd}$- and $3^{rd}$-order flow planes we can also study the correlations between the other order flow planes. These studies have recently gained a lot of 
attention in the field~\cite{Bilandzic:2012an, Bhalerao:2011yg, Jia:2012sa, Qiu:2012uy, Bhalerao:2013ina}.
It was proposed to measured these correlations using the multi-particle mixed harmonic correlations~\cite{Bhalerao:2011yg}:
\begin{multline}
\langle \cos(n_{1}\varphi_{1} + ... + n_{k}\varphi_{k}) \rangle \\
=  \langle v_{n_{1}}...v_{n_{k}} \cos(n_{1}\Psi_{n_{1}} + ... + n_{k}\Psi_{n_{k}}) \rangle .\\
\label{Eq:PPCorrelations}
\end{multline}
These new observables have been measured with 
$v_{3}^{2}\{\Psi_{2}\}$ in ALICE~\cite{ALICE:2011ab} 
and more detailed studies have been presented in a recent work~\cite{Bilandzic:2012an}. 
Hydrodynamic calculations~\cite{Qiu:2012uy} predict that the correlation strength is sensitive to both the initial conditions and the details of the expansion of the system. 
AMPT simulations, which also provide information of initial  and final state,
will help us to understand the role of these correlations in a cascade model.

Using Eq.~(\ref{Eq:PPCorrelations}), we can study the $n^{\rm{th}}$ and $m^{\rm{th}}$ order flow plane correlations in the final state. 
For example, the correlations between ($\Psi_{4}$,$\Psi_{2}$) and ($\Psi_{6}$,$\Psi_{3}$) can be evaluated via: 
\begin{equation}
\begin{aligned}
\langle \cos(4\varphi_{1} - 2\varphi_{2} - 2\varphi_{3}) \rangle &=  \langle v_{4}v_{2}^{2}  \cos(4\Psi_{4} - 4\Psi_{2})\rangle, \\
\langle \cos(6\varphi_{1} - 3\varphi_{2} - 3\varphi_{3}) \rangle &=  \langle v_{6}v_{3}^{2} \langle \cos(6\Psi_{6} - 6\Psi_{3})\rangle. 
\label{Eq:PsiCorrelations}
\end{aligned}
\end{equation}
As discussed in~\cite{Ante-Thesis}, these observables can be directly calculated in terms of a three-particle cumulant:
\begin{equation}
\begin{split}
{\rm QC} \{3\}_{2n, -n, -n}  = & \, \langle \cos [ n (2\varphi_{1} - \varphi_{2} - \varphi_{3}) ] \rangle\\
 =  & \, [ Q_{2n} Q_{n}^{*} Q_{n}^{*} - 2 \cdot |Q_{n}|^{2} - |Q_{2n}|^{2} + 2M ]  \\
 &  \,/ [M(M-1)(M-2)].
\end{split}
\end{equation}
For $n=$2, we get ${\rm QC}\{3\}_{4, -2, -2}$ which is sensitive to the correlations between ($\Psi_{4}, \Psi_{2}$), and analogously for $n =$ 3 which is sensitive to the correlations between ($\Psi_{6}, \Psi_{3}$).

In AMPT, we can also study correlations between the $n^{\rm{th}}$- and $m^{\rm{th}}$-order symmetry planes, in the initial state ($\Phi_{n}$, $\Phi_{m}$).
In this model, we can calculate $\Phi_{n}$ assuming that the initial spatial energy distribution is proportional to the initial spatial parton distribution. Thus, we have:
\begin{equation}
\Phi_{n} = \frac{1}{n}\mathrm{ATan2}\left(\frac{\langle \sin(n\phi) \rangle} {\langle \cos(n\phi)\rangle}+\mathrm{\pi}\right),
\label{Eq:participantplane}
\end{equation}
where $\phi$ is azimuthal angle of the initial partons.

For the ($\Phi_{n}$, $\Phi_{m}$) correlation, we can use, similar to Eq.~(\ref{Eq:PsiCorrelations}): 
\begin{equation}
\begin{aligned}
\langle \cos(4\phi_{1} - 2\phi_{2} - 2\phi_{3}) \rangle = \langle \varepsilon_{4}\varepsilon_{2}^{2} \cos(4\Phi_{4} - 4\Phi_{2})\rangle, \\
\langle \cos(6\phi_{1} - 3\phi_{2} - 3\phi_{3}) \rangle =  \langle \varepsilon_{6}\varepsilon_{3}^{2} \cos(6\Phi_{6} - 6\Phi_{3})\rangle. 
\label{Eq:PhiCorrelations}
\end{aligned}
\end{equation}
If the initial state symmetry plane $\Phi_{n}$ coincides with the flow plane $\Psi_{n}$, the initial ($\Phi_{n}$, $\Phi_{m}$) correlation and the final ($\Psi_{n}$, $\Psi_{m}$) correlation should show a similar centrality dependence, and at least should have the same sign.

We see in Fig.~\ref{fig:7} (top) that the ($\Psi_{4}$, $\Psi_{2}$) correlation has a positive sign, which increases as the centrality increases.
In contrast, the ($\Phi_{4}$, $\Phi_{2}$) correlation is negative and decreases with increasing centrality, plotted in Fig. 4 (bottom).
The negative initial ($\Phi_{4}$, $\Phi_{2}$) correlation and positive final ($\Psi_{4}$, $\Psi_{2}$) correlation observed in the AMPT model are in qualitative agreement with viscous hydrodynamic calculations~\cite{Qiu:2012uy}. There is a clear sign change of the $4^{\rm{th}}$-order and $2^{\rm{nd}}$-order plane correlation during the collision system evolution, both in the transport model and in the hydrodynamic calculations~\cite{Qiu:2012uy}. 
On the other hand, the $6^{\rm{th}}$-order and $3^{\rm{rd}}$-order plane correlation, 
has a negative value in the initial ($\Phi_{6}$, $\Phi_{3}$) correlations 
while it is consistent with 0 within uncertainty in the final state ($\Psi_{6}$, $\Psi_{3}$) correlations.
In addition, we can  see in Fig. 4 (top) that there is no clear difference of the QC$\{3\}$ from all charged particles (solid symbols) and the like-sign (LS) particles (open symbols), which indicates that non-flow contributions to the above observables are small and the sign change of the plane correlations can not be explained by the possible non-flow contributions.

Considering that the initial anisotropy $\varepsilon_{n}$ and the final anisotropic flow $v_{n}$ are both positive, the above results can only be explained if the sign of the genuine correlation, the cosine component, changes during the evolution of the produced system. 
In fact, hydrodynamic calculations suggest that the final $n^{th}$-order flow plane $\Psi_{n}$ might be not only driven by $\Phi_{n}$, but might also have contributions from other symmetry plane(s). The $4^{th}$-order harmonic $v_{4}$ and its associated flow plane $\Psi_{4}$ are determined by a linear and a quadratic response~\cite{Teaney:2012ke},
\begin{equation}
v_{4} \, e^{-i \,4 \, \Psi_{4}} = w_{4} \, e^{-i \, 4 \, \Phi_{4}} + w_{4 \, (22)} \, e^{-i \, 4 \, \Phi_{2}}
\end{equation}
where $w_{4}$ describes the linear response 
, and $w_{4 \, (22)}$ quantifies the non-linear response. 
This nonlinear response in hydrodynamic calculations couples $v_{4}$ to $(v_2)^{2}$ and also couples $v_{6}$ to $v_{3}^{2}$~\cite{Teaney:2012ke}, and clearly both $\Phi_{2}$ and $\Phi_{4}$ contribute to $\Psi_{4}$.
Therefore, these results show that the initial symmetry plane $\Phi_{n}$ and the flow plane $\Psi_{n}$ do not coincident in the AMPT model. Similar results were also observed in the previous transport model calculations using UrQMD model~\cite{Petersen:2010cw}, and confirmed by hydrodynamic calculations~\cite{Qiu:2011iv}.
Thus, the assumption $\Phi_{n}=\Psi_{n}$ used in the so-called true flow calculations (or symmetry plane flow), $v_{n}\{\Phi_{n}\}$~\cite{Alver:2010gr, Ma:2010dv, Ma:2014xfa} is not valid.
In addition, we notice that a new method, named scalar product method~\cite{Bhalerao:2013ina}, is proposed to measure the flow plane correlations without contributions of anisotropic flow. However, it is important to emphasize that this method is based on the assumption that $\langle v_{n}^{2} v_{2n} \rangle =  \langle v_{n}^{2} \rangle \sqrt{\langle v_{2n}^{2} \rangle}$.
In fact, the recent `standard candle' calculations $SC(m,n,-m,-n)_{v}$~\cite{Bilandzic:2013kga, Bhalerao:2014xra} have shown that there are strong (anti-)correlations between event-by-event fluctuations of $v_{n}$ and $v_{m}$, therefore the equality $\langle v_{n}^{2} v_{2n} \rangle =  \langle v_{n}^{2} \rangle \sqrt{\langle v_{2n}^{2} \rangle}$ assumed in the scalar product method does not hold.
Thus, in this paper, we discuss the symmetry plane correlations using only the mixed harmonic correlations.

\begin{figure}[thb]
 \begin{center}
   \includegraphics[width=0.47\textwidth]{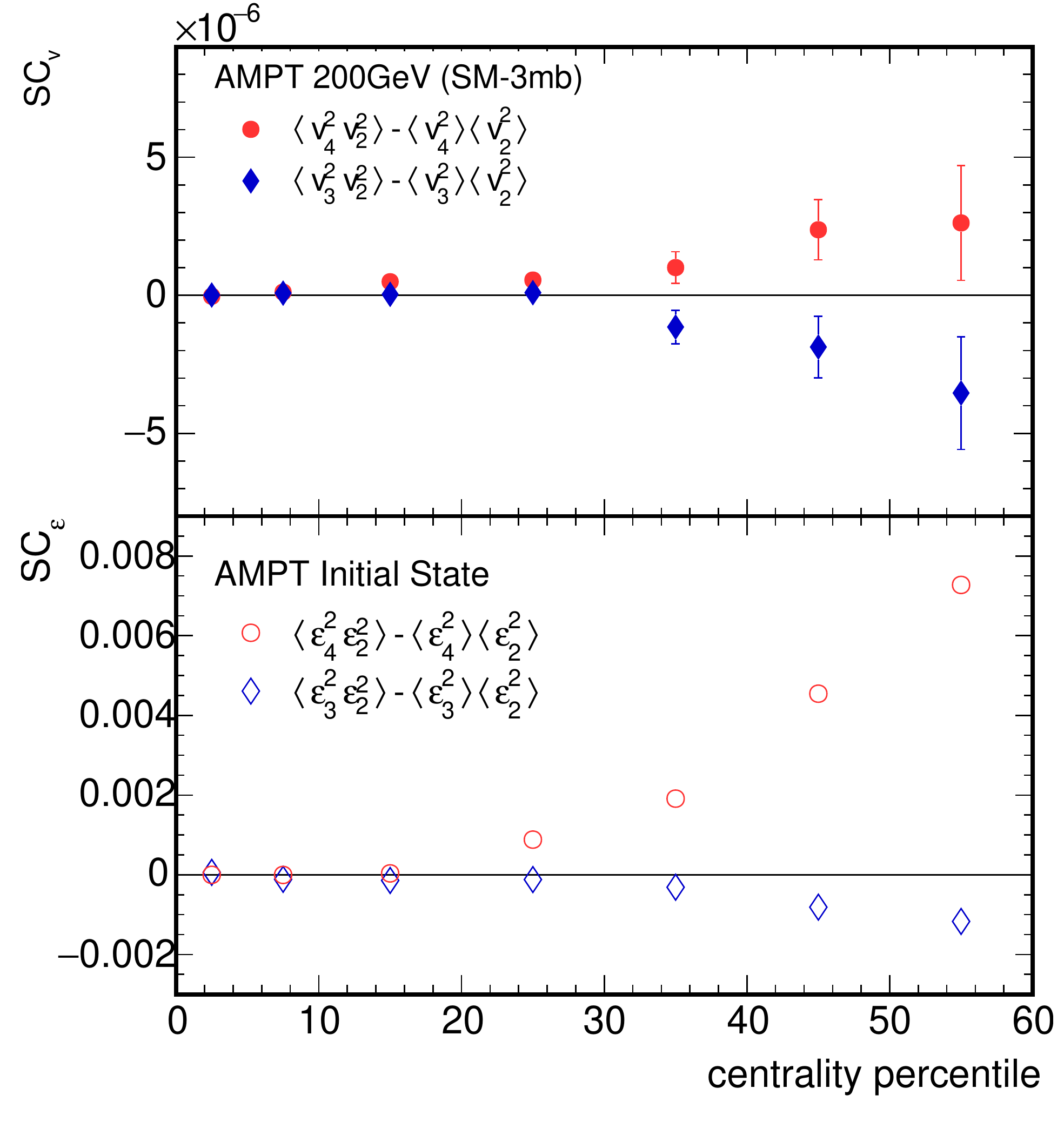}
   \caption{(Color online) Centrality dependence of correlations between the $n^{th}$- and $m^{th}$-order harmonics in the final state (top) and initial state (bottom) in Au+Au collisions at 200 GeV by AMPT StringMelting.  }
 \label{fig:8}
 \end{center}
\end{figure}

Not only the correlations between symmetry planes but also the correlations between different flow harmonics can be investigated via the multi-particle correlation technique. The observable $SC(m,n,-m,-n)_{v}$ was proposed as an unique tool to probe the correlations between different orders of flow harmonics and by design independent of the symmetry planes~\cite{Bilandzic:2013kga}. 

In Fig.~\ref{fig:8} (top) we see a clear non-zero value for both $SC(4,2,-4,-2)_{v}$ (red markers) and $SC(3,2,-3,-2)_{v}$ (blue markers) in the final state. The positive results of $SC(4,2,-4,-2)_{v}$ and negative results of $SC(3,2,-3,-2)_{v}$ are observed for the presented centrality classes. These (anti-)correlations are more pronounced in peripheral collisions. It indicates that finding $v_{2}$ larger than $\overline{v}_{2}$ in an event enhances the probability of finding $v_{4}$ larger than $\overline{v}_{4}$ and, in addition, the probability of finding $v_{3}$ smaller than $\overline{v}_{3}$ in that event, as was shown in a previous study~\cite{Bilandzic:2013kga}. Also we investigate the correlations of the $n^{th}$- and $m^{th}$-order harmonics in the initial state, calculating:
\begin{equation}
SC(m,n,-m,-n)_{\varepsilon} = \langle \varepsilon_{m}^{2}  \, \varepsilon_{n}^{2} \rangle - \langle \varepsilon_{m}^{2} \rangle \langle \varepsilon_{n}^{2} \rangle 
\end{equation}
The result is presented in Fig.~\ref{fig:8} (bottom). We see a positive and increasing trend for $SC(4,2,-4,-2)_{\varepsilon}$ while a negative and decreasing trend of $SC(3,2,-3,-2)_{\varepsilon}$ is observed. They capture the rough trend of $SC(4,2,-4,-2)_{v}$ and $SC(3,2,-3,-2)_{v}$ in the final state but cannot quantitatively describe the centrality dependence. This suggests indeed a correlation between the initial state $SC(m,n,-m,-n)_{\varepsilon}$ and final state $SC(m,n,-m,-n)_{v}$, but this might not be the only contribution to the final state. In addition the system evolution, might also modify the strength of $SC(m,n,-m,-n)_{v}$. 
In a previous study~\cite{Bilandzic:2013kga}, three configurations of the AMPT model, named: (a) 3mb; (b) 10mb; and (c)  10mb no rescattering, have been investigated to better understand the $SC(m,n,-m,-n)_{v}$ calculations. In general, the configuration of (a) 3mb generates weaker partonic interactions during the system evolution compared to (b) 10mb. And the hadronic interactions are turned off for (c) 10mb no rescattering. The comparisons between these three configurations shows that the value of $SC(m,n,-m,-n)_{v}$ depends on both the partonic and the hadronic interactions. These results suggest that the sign of $SC(m,n,-m,-n)_{v}$ is determined by the initial state $SC(m,n,-m,-n)_{\varepsilon}$ and the magnitude is modified by the multiple interactions during the system evolution.  

As we discussed above, the Elliptic-Power function gives a better description of the $p.d.f.$ of each single harmonic. However, it is an open question at the moment how the joint underlying $p.d.f.$ including different order symmetry planes and harmonics is described and, additionally, if these correlations between different order symmetry planes and harmonics modify the single harmonics $p.d.f.$ The investigations presented here begins to answer these open questions. Nevertheless, many more measurements between different order symmetry planes and harmonics are necessary to reasonably constrain a joint $p.d.f.$, and ultimately lead to new insights into the nature of the fluctuation of the created matter in heavy ion collisions.

\section{Summary}
In this paper we presented the calculations of the initial and final state anisotropies in Au+Au collisions at $\sqrt{s_{\rm NN}}=200$ GeV using AMPT simulations. 
It is found that the Elliptic-Power function is the only $p.d.f.$ so far which describes the event-by-event distributions of the eccentricity as well as the triangularity, the anisotropic flow from multi-particle cumulants and the relative flow fluctuations very well. 
In addition, the correlations between different order symmetry planes and harmonics have been 
investigated. A different centrality dependence of these correlations in the initial and final state has been observed within the same framework of a transport model. This result indicates that both the fluctuations in the initial geometry and the dynamical evolution of the medium in the final state plays an important roles. 
It is currently still unclear how well the underlying joint $p.d.f.$ matches the experimental data as these predictions still have to be tested at RHIC and at the LHC.
The study presented in this paper should help us better understand the fluctuations of created matter in heavy ion experiments.

\section{Acknowledgments}
The authors thanks A.\,Biland\v{z}i\'{c}, R.\,Bertens and K.\,Gulbrandsen for the comments on the manuscripts, thanks P.\,Christakoglou, A.\,Tang, S.\,Voloshin, F.Q.\,Wang for fruitful discussion. Part of this work is inspired by the discussion with J.\,Y.\,Ollitrault.
The authors are supported by the Danish Council for Independent Research, Natural Sciences, and the Danish National Research Foundation (Danmarks Grundforskningsfond),
MOST of China under 973 Grant 2015CB856901, the National Natural Science Foundation of China (NSFC) under Grants 11135011, 11228513, 11221504, FOM and NWO of the Netherlands.

%

%
\end{document}